\documentstyle[twoside,epsfig]{article}

\catcode`\@=11
\long\def\@makefntext#1{
\protect\noindent \hbox to 3.2pt {\hskip-.9pt
$^{{\eightrm\@thefnmark}}$\hfil}#1\hfill}		

\def\thefootnote{\fnsymbol{footnote}}
\def\@makefnmark{\hbox to 0pt{$^{\@thefnmark}$\hss}}	

\def\ps@myheadings{\let\@mkboth\@gobbletwo
\def\@oddhead{\hbox{}
\rightmark\hfil\eightrm\thepage}
\def\@oddfoot{}\def\@evenhead{\eightrm\thepage\hfil
\leftmark\hbox{}}\def\@evenfoot{}
\def\sectionmark##1{}\def\subsectionmark##1{}}

\oddsidemargin=\evensidemargin
\renewcommand{\thefootnote}{\fnsymbol{footnote}}

\newcounter{sectionc}
\newcounter{subsectionc}
\newcounter{subsubsectionc}
\renewcommand{\section}[1] {\vspace{12pt}\addtocounter{sectionc}{1}
\setcounter{subsectionc}{0}\setcounter{subsubsectionc}{0}\noindent
	{\tenbf\thesectionc. #1}\par\vspace{5pt}}
\renewcommand{\subsection}[1] {\vspace{12pt}
\addtocounter{subsectionc}{1}\setcounter{subsubsectionc}{0}\noindent
	{\bf\thesectionc.\thesubsectionc.
        {\kern1pt \bfit #1}}\par\vspace{5pt}}
\renewcommand{\subsubsection}[1] {\vspace{12pt}
\addtocounter{subsubsectionc}{1}\noindent
        {\tenrm\thesectionc.\thesubsectionc.\thesubsubsectionc.
	{\kern1pt \tenit #1}}\par\vspace{5pt}}

\newcounter{appendixc}
\newcounter{subappendixc}[appendixc]
\newcounter{subsubappendixc}[subappendixc]
\renewcommand{\thesubappendixc}{\Alph{appendixc}.
        \arabic{subappendixc}}
\renewcommand{\thesubsubappendixc}{\Alph{appendixc}.
        \arabic{subappendixc}.\arabic{subsubappendixc}}

\renewcommand{\appendix}[1] {\vspace{12pt}
        \refstepcounter{appendixc}
        \setcounter{figure}{0}
        \setcounter{table}{0}
        \setcounter{lemma}{0}
        \setcounter{theorem}{0}
        \setcounter{corollary}{0}
        \setcounter{definition}{0}
        \setcounter{equation}{0}
        \renewcommand{\thefigure}{\Alph{appendixc}.\arabic{figure}}
        \renewcommand{\thetable}{\Alph{appendixc}.\arabic{table}}
        \renewcommand{\theappendixc}{\Alph{appendixc}}
        \renewcommand{\thelemma}{\Alph{appendixc}.\arabic{lemma}}
        \renewcommand{\thetheorem}{\Alph{appendixc}.\arabic{theorem}}
        \renewcommand{\thedefinition}{\Alph{appendixc}.
         \arabic{definition}}
        \renewcommand{\thecorollary}{\Alph{appendixc}.
         \arabic{corollary}}
        \renewcommand{\theequation}{\Alph{appendixc}.
         \arabic{equation}}
        \noindent{\tenbf Appendix \theappendixc #1}\par\vspace{5pt}}
\newcommand{\subappendix}[1] {\vspace{12pt}
        \refstepcounter{subappendixc}
        \noindent{\bf Appendix \thesubappendixc. {\kern1pt \bfit #1}}
	\par\vspace{5pt}}
\newcommand{\subsubappendix}[1] {\vspace{12pt}
        \refstepcounter{subsubappendixc}
        \noindent{\rm Appendix \thesubsubappendixc.
        {\kern1pt \tenit #1}}\par\vspace{5pt}}

\newcommand{\textlineskip}{\baselineskip=13pt}
\newcommand{\smalllineskip}{\baselineskip=10pt}

\def\eightcirc{
\begin{picture}(0,0)
\put(4.4,1.8){\circle{6.5}}
\end{picture}}
\def\eightcopyright{\eightcirc\kern2.7pt\hbox{\eightrm c}}

\newcommand{\pub}[1]{{\begin{center}\footnotesize\smalllineskip
	Preprint No. #1\\
	\end{center}
	}}

\def\abstracts#1#2#3{{
	\centering{\begin{minipage}{4.5in}\baselineskip=10pt
        \footnotesize
	\parindent=0pt #1\par
	\parindent=15pt #2\par
	\parindent=15pt #3
	\end{minipage}}\par}}

\renewenvironment{thebibliography}[1]
	{\frenchspacing
	 \ninerm\baselineskip=11pt
	 \begin{list}{\arabic{enumi}.}
	{\usecounter{enumi}\setlength{\parsep}{0pt}
	 \setlength{\leftmargin 12.7pt}{\rightmargin 0pt}
	 \setlength{\itemsep}{0pt} \settowidth
	{\labelwidth}{#1.}\sloppy}}{\end{list}}

\newcounter{itemlistc}
\newcounter{romanlistc}
\newcounter{alphlistc}
\newcounter{arabiclistc}

\newcommand{\fcaption}[1]{
        \refstepcounter{figure}
        \setbox\@tempboxa = \hbox{\footnotesize Fig.~\thefigure. #1}
        \ifdim \wd\@tempboxa > 5in
           {\begin{center}
        \parbox{5in}{\footnotesize\smalllineskip Fig.~\thefigure. #1}
            \end{center}}
        \else
             {\begin{center}
             {\footnotesize Fig.~\thefigure. #1}
              \end{center}}
        \fi}

\newcommand{\tcaption}[1]{
        \refstepcounter{table}
        \setbox\@tempboxa = \hbox{\footnotesize Table~\thetable. #1}
        \ifdim \wd\@tempboxa > 5in
           {\begin{center}
        \parbox{5in}{\footnotesize\smalllineskip Table~\thetable. #1}
            \end{center}}
        \else
             {\begin{center}
             {\footnotesize Table~\thetable. #1}
              \end{center}}
        \fi}

\def\@citex[#1]#2{\if@filesw\immediate\write\@auxout
	{\string\citation{#2}}\fi
\def\@citea{}\@cite{\@for\@citeb:=#2\do
	{\@citea\def\@citea{,}\@ifundefined
	{b@\@citeb}{{\bf ?}\@warning
	{Citation `\@citeb' on page \thepage \space undefined}}
	{\csname b@\@citeb\endcsname}}}{#1}}

\newif\if@cghi
\def\cite{\@cghitrue\@ifnextchar [{\@tempswatrue
	\@citex}{\@tempswafalse\@citex[]}}
\def\citelow{\@cghifalse\@ifnextchar [{\@tempswatrue
	\@citex}{\@tempswafalse\@citex[]}}
\def\@cite#1#2{{$\null^{#1}$\if@tempswa\typeout
	{IJCGA warning: optional citation argument
	ignored: `#2'} \fi}}

\def\pmb#1{\setbox0=\hbox{#1}
	\kern-.025em\copy0\kern-\wd0
	\kern.05em\copy0\kern-\wd0
	\kern-.025em\raise.0433em\box0}

\def\fnt#1#2{\footnotetext{\kern-.3em
	{$^{\mbox{\scriptsize #1}}$}{#2}}}

\def\fpage#1{\begingroup
\voffset=.3in
\thispagestyle{empty}\begin{table}[b]\centerline{\footnotesize #1}
	\end{table}\endgroup}

\font\tenrm=cmr10
\font\tenit=cmti10
\font\tenbf=cmbx10
\font\bfit=cmbxti10 at 10pt
\font\ninerm=cmr9

\font\eightrm=cmr8

\def\qed{\hbox{${\vcenter{\vbox{		
   \hrule height 0.4pt\hbox{\vrule width 0.4pt height 6pt
   \kern5pt\vrule width 0.4pt}\hrule height 0.4pt}}}$}}

\renewcommand{\thefootnote}{\fnsymbol{footnote}}

\input epsf
\global\arraycolsep=2pt
\def\spose#1{\hbox to 0pt{#1\hss}}
\def\lsim{\mathrel{\spose{\lower 3pt\hbox{$\mathchar"218$}}
 \raise 2.0pt\hbox{$\mathchar"13C$}}}
\def\gsim{\mathrel{\spose{\lower 3pt\hbox{$\mathchar"218$}}
 \raise 2.0pt\hbox{$\mathchar"13E$}}}

\renewcommand{\theequation}{\thesection.\arabic{equation}}
\def\laq{\raise 0.4ex\hbox{$<$}\kern -0.8em\lower 0.62
ex\hbox{$\sim$}}
\def\gaq{\raise 0.4ex\hbox{$>$}\kern -0.7em\lower 0.62
ex\hbox{$\sim$}}

\def\beq{\begin{equation}}
\def\eeq{\end{equation}}
\def\bea{\begin{eqnarray}}
\def\eea{\end{eqnarray}}

\def \pa {\partial}
\def \ra {\rightarrow}

\def \Da {\Delta}
\def \b {\beta}
\def \a {\alpha}

\def \sg {\sigma}
\def \da {\delta}
\def \ep {\epsilon}
\def \r {\rho}
\def \om {\omega}
\def \Om {\Omega}
\def \noi {\noindent}

\begin{document}

\begin{titlepage}

\begin{flushright}
DFTT-32/98\\
gr-qc/9806073
\end{flushright}

\vspace{2 cm}

\begin{center}
\Large\bf Weighing the String Mass \\
with the COBE Data
\end{center}

\vspace{1.5cm}

\begin{center}
M. Gasperini\\
{\sl Dipartimento di Fisica Teorica, Universit\`a di Torino,}\\
{\sl Via P. Giuria 1, 10125 Turin, Italy}\\
and\\
{\sl Istituto Nazionale di Fisica Nucleare, Sezione di Torino, Turin, Italy}\\
\end{center}

\vspace{1.5cm}

\begin{abstract}
\noi
In the context of the pre-big bang scenario the large-scale CMB 
anisotropy can be seeded by a primordial background of very light (or 
massless) axion fluctuations. In that case the slope of the temperature 
anisotropy spectrum, allowed by present observations, defines an allowed 
range of values for the string mass scale. Conversely, from the 
theoretical expected value of the string scale we can predict the slope 
of the anisotropy spectrum. In both cases there is a remarkable 
agreement between observations and theoretical expectations. 

\end{abstract}

\vspace{1.5cm}
\begin{center}
------------------------------

\vspace{1.5cm}
To appear in \\
{\sl Proc. of the Euroconference ``Fifth Paris Cosmology 
Colloquium"}\\
Observatoire de Paris, 3-5 June 1998 -- 
Eds. H. J. De Vega and N. Sanchez\\
(World Scientific, Singapore)
\end{center}
 \vspace{1.5cm}
\vfill

\end{titlepage}


\normalsize\textlineskip
\thispagestyle{empty}
\setcounter{page}{1}


\vspace*{0.11truein}

\fpage{1}

\centerline{\bf WEIGHING THE STRING MASS WITH THE COBE DATA}
\vspace*{0.27truein}

\centerline{\footnotesize MAURIZIO GASPERINI}
\vspace*{0.015truein}
\centerline{\footnotesize\it Dipartimento di Fisica Teorica, 
Universit\`a di Torino,}
\baselineskip=10pt
\centerline{\footnotesize  {\it Via P. Giuria 1, 10125, Turin, Italy}}
\baselineskip=10pt
\centerline{\footnotesize and {\it Istituto Nazionale di Fisica Nucleare,
Sezione di Torino, Turin, Italy}}

\vspace*{0.3truein}
\abstracts
{In the context of the pre-big bang scenario the large-scale CMB 
anisotropy can be seeded by a primordial background of very light (or 
massless) axion fluctuations. In that case the slope of the temperature 
anisotropy spectrum, allowed by present observations, defines an allowed 
range of values for the string mass scale. Conversely, from the 
theoretical expected value of the string scale we can predict the slope 
of the anisotropy spectrum. In both cases there is a remarkable 
agreement between observations and theoretical expectations. }
{}{}
\vspace*{0.225truein}
\pub{DFTT-32/98;~~~~~~~~~~ E-print Archives: gr-qc/9806073}
\vspace*{0.8pt}\textlineskip

\textheight=7.8truein
\setcounter{footnote}{0}
\renewcommand{\thefootnote}{\alph{footnote}}

\vspace*{0.125truein}

\renewcommand{\theequation}{1.\arabic{equation}}
\setcounter{equation}{0}
\section{Introduction}
\label{sec:1}
\noindent
The aim of this paper is to review, and briefly discuss, 
a possible mechanism for generating the large-scale CMB anisotropy, 
based on a primordial background of axion fluctuations acting as 
seeds for scalar metric perturbations\cite{1,2,3}.

Such a mechanism is particularly appropriate 
to pre-big bang models\cite{4} 
formulated in a string cosmology context, since in that case it seems 
difficult\cite{5} to generate the observed anisotropy through the 
standard inflationary mechanism. Let me explain why. 

At very large angular scales, the temperature anisotropy spectrum is 
determined by the metric fluctuation spectrum $\Phi_k$ through the
well-know  Sachs-Wolfe (SW) effect\cite{6}: 
\beq 
\left(\Da T\over T\right)_k \sim \Phi_k .
\label{11}
\eeq
Metric 
fluctuations, directly amplified by the accelerated evolution of the 
background, have a spectrum that depends on the value of the  
Hubble scale at the time of horizon crossing,
\beq
\Phi_k \sim \left(H\over M_p\right)_k
\label{12}
\eeq
($M_p$ is the Planck mass). In the standard de Sitter (or quasi-De 
Sitter) inflationary scenario $H$ is constant in time, so that the 
spectrum is scale invariant. A typical normalization of the 
spectrum, corresponding to inflation occurring roughly at the GUT
scale,  
\beq 
{H\over M_p} \sim {{\rm GUT~curvature~scale \over PLANCK~scale}}
\sim 10^{-5}, 
\label{13}
\eeq
is  thus perfectly consistent with the anisotropy observed 
at the present horizon scale, $\Da T/T \sim 10^{-5}$, and with the fact 
that the spectrum is scale-invariant. 

Why this simple mechanism does not work in a string cosmology 
context? In string cosmology models the curvature scale grows with
time, so that  the spectrum of metric fluctuations (\ref{12}) grows
with frequency.  In addition, the natural inflation scale corresponds to
the string   scale, so that the normalization of the spectrum, at the 
end-point frequency $k_1$, is controlled by the ratio
\beq 
\left(H\over M_p\right)_{k_1} 
\sim {{\rm STRING~curvature~scale \over PLANCK~scale}} \sim
10^{-2}.
\label{14}
\eeq

We are thus led to the situation qualitatively illustrated in Fig. 1. 
For pre-big bang models the slope of the spectrum is too steep,  and
the  normalization too high, to be compatible with COBE
observations\cite{7}.  The slope  is so steep, however, that the
contribution of metric  fluctuations to $\Da T/T$ is certainly negligible
at the COBE scale. So,  on one hand there is no contradiction with
observations, namely the COBE  data cannot be used to rule out
pre-big bang models. On the other hand,  the problem remains: how to
explain the observed anisotropy if the  contribution of metric
fluctuations is so small? 

\begin{figure}[htb]
\vspace{10cm}
\includegraphics{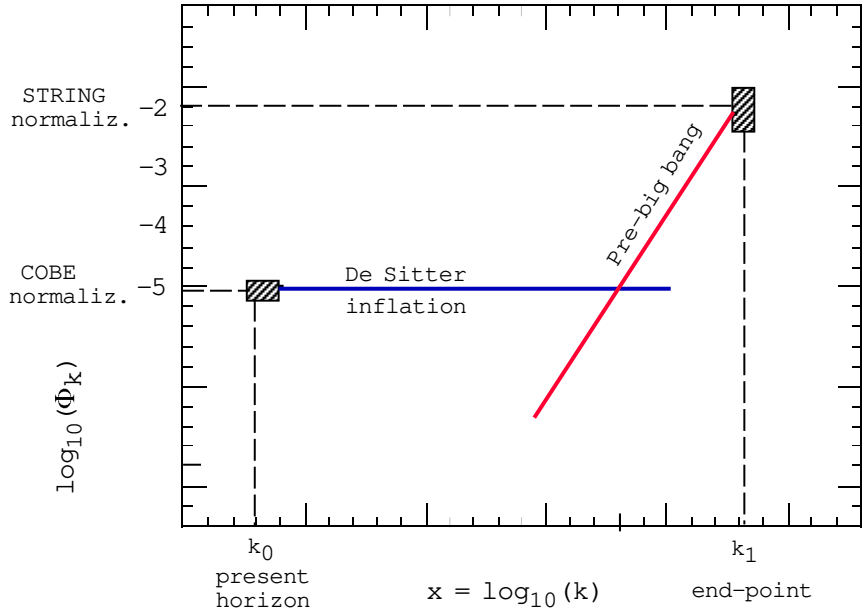}
{\caption{\label{fig:f1}
{\sl The contribution of primordial metric fluctuations to $\Da T/
T$, in the pre-big bang scenario, is expected to be negligible at
the COBE scale. }}}
\end{figure}

A possible answer to this question comes from the observation  that
the  previous argument applies to the primordial spectrum of metric 
fluctuations, {\em directly} amplified by the accelerated 
evolution of the background. There is an additional indirect 
contribution to the final metric perturbation spectrum, however, 
arising  from the quantum fluctuations of other fields (let me call
them,  generically, $\sg$), amplified during inflation. Even if such 
fluctuations are eventually negligible as  sources of the metric 
background, $\r_\sg \ll \r_c$, their inhomogeneous stress tensor 
generates metric fluctuations according to the standard gravitational 
equations, and they can act as ``seeds" for temperature anisotropies 
through the SW effect, as before:
\beq
{\r_\sg \over \r_c} \sim \Phi \sim {\Da T\over T}
\label{15}
\eeq

Why the seed mechanism can work? First of all because, unlike metric 
perturbations, there are fields whose fluctuations can be  amplified
with  a flat spectrum even in the context of the pre-big bang
scenario. 

Second because the contribution to $\Da T/T$ is quadratic in the seed 
fields, and not linear like in case of metric perturbations. So, even if 
the amplitude of seed fluctuations is still normalized at the string
curvature scale, the square of the amplitude is
not very far  from the expected value $10^{-5}$:
\beq 
{\Da T \over T} \sim \Phi \sim \sg^2 
\sim \left({\rm STRING~curvature~scale \over PLANCK~scale}\right)^2
\sim 10^{-4}.
\label{16}
\eeq
In addition, we must recall that the string normalization is      
imposed at  the end-point of the spectrum\cite{8}  (roughly, at the
GHz scale), while  COBE observations constrain the spectrum at the
present horizon scale  ($\sim 10^{-18}$Hz). A very small (blue) tilt of
the seed  field spectrum  is thus enough to make compatible the COBE
normalization and the string  normalization, as illustrated in Fig. 2. 

\begin{figure}[htb]
\vspace{10cm}
\includegraphics{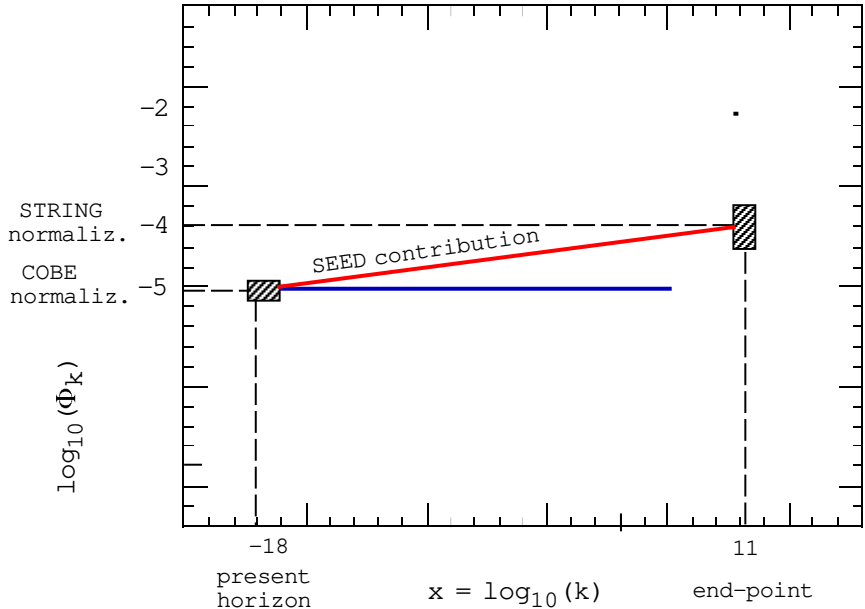}
{\caption{\label{fig:f2}
{\sl The amplitude of metric fluctuations induced by seeds may be 
consistent both with the COBE and the string normalization of the 
spectrum. }}}
\end{figure}

The basic question now becomes: are there fields, in the context  of
the  pre-big bang scenario, whose fluctuations can be amplified with a
flat  enough spectrum, so as to seed metric fluctuations and to fit 
consistently the observed anisotropy? 

In the following Sections I will present two possible examples that 
seem to be promising: the case of massless and massive axion 
fluctuations.

\vskip 1 cm
\renewcommand{\theequation}{2.\arabic{equation}}
\setcounter{equation}{0}
\section{Massless axions as seeds of large-scale anisotropy}
\label{sec:2}
\noindent

A first possible candidate for seeding the large-scale anisotropy is a 
stochastic background of massless pseudoscalar fluctuations\cite{9}. 
I will 
take, as a particular example, the so-called ``universal" axion of 
string theory, namely the four-dimensional  dual $\sg$ of the
Kalb-Ramond  antisymmetric tensor $H_{\mu\nu\a}$, appearing in the
low-energy string  effective action:
\bea
&&
S=-\int d^4x \sqrt{-g} e^{-\phi}\left[R+(\pa_\mu \phi)^2 -{1\over 12} 
H_{\mu\nu\a}^2 \right], \nonumber\\
&&
H^{\mu\nu\a} =e^{\phi} \ep^{\mu\nu\a\b} \pa_\b \sg .
\label{21}
\eea
The whole discussion can be applied, however, to any type of
pseudoscalar fluctuation amplified with a flat enough primordial
energy  spectrum. 

I will concentrate my discussion on three points. First, I have to show 
that such axion fluctuations can be amplified with a final 
scale-invariant distribution of their spectral energy density 
$\Om_\sg$: 
\beq
\Om_\sg(k,\eta) = {d\r_\sg(k,\eta)\over \r_c d \ln k} \sim 
{\rm scale~invariant}.
\label{22}
\eeq
Second, I will show that the scalar metric fluctuation on a given scale
$k$, at the  time the scale re-enters the horizon, is precisely
determined by the axion  energy distribution evaluated at the
conformal time of re-entry,  $\eta_{re} \simeq k^{-1}$:
\beq
\Phi_k(\eta_{re}) \sim \Om_\sg (k, \eta_{re}). 
\label{23}
\eeq
Third, I will show that the dominant contribution  to the SW effect
comes  from a scale at the time it re-enters the horizon, so that the
final  temperature spectrum exactly reproduces the primordial seed
spectrum: 
\beq
\left(\Da T\over T\right)_k \sim \Phi_k(\eta_{re}) \sim \Om_\sg (k, 
\eta_{re}). 
\label{24}
\eeq
These last two results are far from being trivial, being a consequence 
of the particular time-dependence of the Bardeen spectrum  induced
by  axion fluctuations (these results 
do not apply, for instance, to electromagnetic 
fluctuations\cite{1}). I will give in this paper only a sketch of the 
arguments leading to the above results. A detailed derivation can be 
found in Refs. [1,2]. 

1) The possibility of a 
flat axion spectrum\cite{9} can be easily checked 
by considering the axion perturbation equation written in terms of the 
canonical variable, $\psi=\sg \xi$, and of the ``pump" field 
$\xi=ae^{\phi/2}$ (here $\phi$ is the dilaton, and $a$ is the 
four-dimensional scale factor of the string frame metric). In the 
conformal time gauge one obtains from eq. (\ref{21}) the effective 
action
\beq
S= \int d^3x d\eta~ a^2 e^\phi \left(\sg^{\prime 2} +\sg 
\nabla^2\sg \right),
\label{25}
\eeq
and the perturbation equation (for the Fourier modes $\psi_k$)
\beq
\psi_k'' +\left[k^2-(\xi''/\xi)\right]\psi_k =0
\label{26}
\eeq
(the prime denotes differentiation with respect to the conformal time 
$\eta$). Assuming, for instance, a power-law evolution of the pre-big 
bang background, $\xi \sim |\eta|^\a$, the perturbation equation 
reduces  to a Bessel equation, and the normalized solution can be
written in  terms of the Hankel functions $H_\nu$ as
\beq
\psi_k = \eta^{1/2} H_\nu^{(2)}(k\eta), ~~~~~~~~~~~~~~~~~
\nu= |\a-1/2|,
\label{26a}
\eeq
where the Bessel index $\nu$ depends on the kinematic of the 
background. The spectral energy density, for modes re-entering in the 
radiation era, depends finally on the background as
\beq
\Om_\sg (k) \sim k^{3-2\nu}.
\label{27}
\eeq

If we take now a very simple, higher-dimensional but isotropic 
vacuum  solution of the string cosmology equations, in $d=3+n$ spatial 
dimensions\cite{10},
\beq
a \sim |\eta |^{-1/(1+\sqrt{d})}, ~~~~~~~~~~~~~~~~~
e^{\phi/2} \sim |\eta |^{-{3+\sqrt d \over 
2(1+\sqrt d)}}, 
\label{28}
\eeq
we find that the spectral index depends on $d$,
\beq
3-2\nu= {\sqrt d -3\over 1+\sqrt d}, 
\label{29}
\eeq
and that the spectrum may be flat ($3-2\nu=0$), in particular\cite{11}, 
for $d=9$. 

It is important to stress that a flat spectrum is possible, in the 
previous background,  for axion 
fluctuations, but impossible for metric fluctuations which are 
characterized by a different pump field\cite{12}, $\xi=a e^{-\phi/2}$. 
With this pump field, in the same background (\ref{28}), both  the
power  $\a$ and the Bessel index $\nu$ are independent of $d$, and
the spectrum  is always growing with a cubic slope, in any number of
dimensions: \beq
\xi = ae^{-\phi/2} \sim |\eta|^{1/2}, ~~~~~~~~~~~~~~~~~~
3-2\nu=3.
\label{210}
\eeq

2) Let us now compute the spectrum of metric perturbations  seeded
by a  flat, primordial distribution of axion fluctuations. Define, as
usual,  the power spectrum of the Bardeen potential, $P_\Phi(k)$, in
terms of  the Fourier transform of the two-point correlation function:
\beq
\int {d^3 k\over (2 \pi k)^3} e^{i {\bf k}\cdot({\bf x}-{\bf x'})}
P_\Phi(k) =\langle \Phi(x) \Phi(x')\rangle
\label{211}
\eeq
(the brackets denote spatial average, or expectation value if 
perturbations are quantized). The square root of the two-point 
function,  evaluated at a comoving distance $k^{-1}$, represents the
typical  amplitude of fluctuations on a scale $k$:
\beq
\left(\langle \Phi(x) \Phi(x')\rangle\right)^{1/2}_{|x-x'|=k^{-1}} 
\sim k^{3/2} |\Phi_k| .
\label{212}
\eeq
Define also the power spectrum of the seed stress tensor, in the same 
way (no sum over $\mu$, $\nu$):
\beq
\int {d^3 k\over (2 \pi k)^3} e^{i {\bf k}\cdot({\bf x}-{\bf x'})}
P_\mu^\nu(k) =\langle T_\mu^\nu(x) T_\mu^\nu(x')\rangle
-\langle T_\mu^\nu (x)\rangle ^2 .
\label{213}
\eeq
Metric fluctuations and seed fluctuations are related by the 
cosmological perturbation equations. By taking into account the 
important contribution of the off-diagonal components of the axion 
stress tensor one finds, typically, that Bardeen spectrum and axion 
energy density spectrum are related by\cite{1}:
\beq
P_\Phi^{1/2}(k) \sim G \left(a \over k\right)^2 P_\r^{1/2}(k),
\label{214}
\eeq
where
\bea
\int {d^3 k\over (2 \pi k)^3} e^{i {\bf k}\cdot({\bf x}-{\bf x'})}
P_\r(k) &=&\langle \r_\sg(x) \r_\sg(x')\rangle
-\langle \r_\sg(x) \rangle^2\nonumber \\
&\sim & \langle \sg^{\prime 2}(x) \sg^{\prime 2}(x')\rangle
-\langle \sg^{\prime 2}(x)\rangle^2 + ...
\label{215}
\eea
It may be interesting to note that the two-point correlation  function
of  the energy density becomes a four-point function of the seed
field,  since the energy is quadratic in the axion field. 

Using the condition of stochastic average for the axion field\cite{1},
\beq
=\langle \sg'({\bf k}, \eta)\sg^{\prime \ast}
({\bf k'}, \eta)\rangle= (2 \pi)^3 \da^3 (k-k') \Sigma(
{\bf k}, \eta), 
\label{216}
\eeq
we find that the energy density spectrum reduces to a convolution of 
Fourier transforms,
\beq
P_\r(k) \sim {k^3\over a^4} \int d^3p~ \Sigma(p) 
\Sigma(|k-p|)+ ... ~~,
\label{217}
\eeq
which is dominated by the region $p\eta \sim 1$ for a flat  enough
axion  spectrum\cite{1,2}. By expressing the convolution through the
spectral  energy density $\Om_\sg$, and evaluating the Bardeen
potential at the  time of re-entry $\eta_{re} \sim k^{-1}$, we are led
finally to relate  the Bardeen spectrum and the axion spectrum as
\beq
P_\Phi^{1/2} (k, \eta_{re}) 
\sim k^{3/2} \left|\Phi_k(\eta_{re}\right| 
\sim \Om_\sg(k, \eta_{re}).
\label{218}
\eeq
In the next (and last) step of my discussion I will explain why 
we are interested in metric fluctuations evaluated at the time of 
re-entry.

3) Let us come back, finally, to the seed contribution to $\Da T
/T$. In the multipole expansion of the temperature anisotropies, 
\beq
\left\langle{\delta T\over T}({\bf
n}){\delta T\over T}({\bf n}') \right\rangle_{{~}_{\!\!({\bf n\cdot
n}'=\cos\vartheta)}} ~~=~~~~
  {1\over 4\pi}\sum_\ell(2\ell+1)C_\ell P_\ell(\cos\vartheta)~,
\label{219}
\eeq
the coefficients $C_\ell$, at very large angular scales 
($\ell \ll 100$), are determined by the SW effect as 
follows\cite{1,13}: 
\beq
C_\ell^{SW} ={2\over \pi}\int
d~(\ln k) \left \langle \left[\int^{k\eta_0}_{k\eta_{dec}} d(k\eta)~
k^{3/2} (\Psi -\Phi)({\bf k},
\eta)j_{\ell}^{\prime}\left(k\eta_0-k\eta \right)\right]^2 
\right\rangle .
\label{220}
 \eeq
Here $\Phi$ and $\Psi$ are the two-independent components of the 
gauge-invariant Bardeen potential, and 
$j_\ell$ are the spherical Bessel functions. Eq. (\ref{220}) takes into 
account both the 
``ordinary" and the ``integrated" SW contribution, namely the 
complete  distortion of the geodesics of the CMB photons (due to
shifts in the  gravitational potential), from the time of decoupling
$\eta_{dec}$ down  to the present time $\eta_0$. By inserting the
Bardeen potential  determined by the axion field, one now finds that
the time integral is  dominated by the region $k\eta \sim 1$. Using eq.
(\ref{218}) we  obtain 
\bea
C_\ell^{SW} &\sim & \int d (\ln k)~ k^{3/2} |\Phi_k(\eta_{re})|^2 
j_\ell (k\eta_0)|^2
\nonumber\\
&\sim & \int d (\ln k)~ \Om_\sg^2 (k, \eta_{re})
\label{221}
\eea
($j_\ell$ are the spherical Bessel functions). 
Here is why it was important to evaluate the Bardeen spectrum 
at the time of re-entry, $\eta_{re} \sim k^{-1}$. 

From the final expression that gives the multipole coefficients 
in terms  of the axion spectral distribution\cite{1,2} 
we can extract, in 
particular, the value of the quadrupole coefficient $C_2$:
\beq
C_2 \simeq \Om_\sg ^2 (k_0, \eta_0) \simeq 
\left(M_s\over M_p\right)^4 \left(k_0\over k_1\right)^{n-1},
\label{222}
\eeq
where $n$ is the spectral index (sufficiently near to $1$) 
characterizing the primordial axion distribution, $k_0$ is the comoving 
scale of the present horizon, and $k_1$ the end-point of the spectrum, 
namely the maximal amplified comoving frequency. 

The peak amplitude of the axion spectrum, at the  end-point
frequency, is  controlled by the fundamental ratio between string and
Planck mass,  $M_s/M_p$. The quadrupole coefficient, on the other
hand, is presently  determined by COBE as\cite{14}
\beq
C_2 = (1.9 \pm 0.23) \times 10^{-10}.
\label{223}
\eeq
This experimental value, inserted into eq. (\ref{222}), implies a 
relation between the string mass and the spectral index of the 
temperature anisotropy, which is illustrated in Fig. 3. 

\begin{figure}[htb]
\epsfxsize=11cm
\centerline{\epsfbox{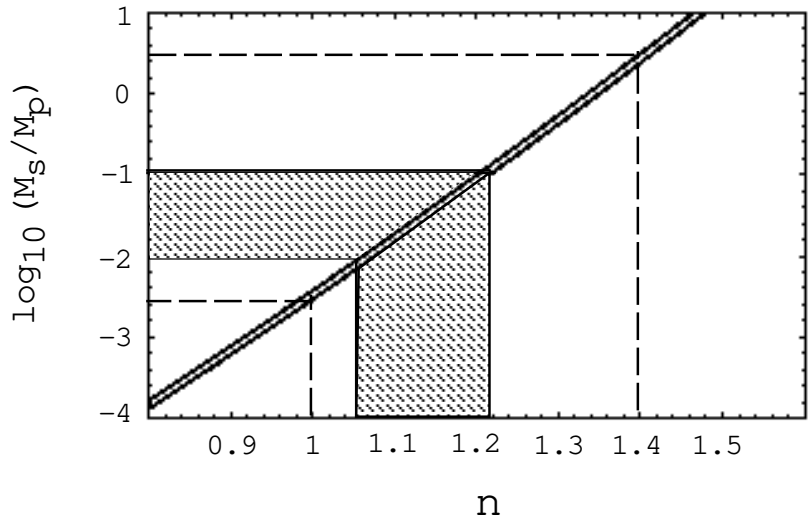}}
\centerline{\parbox{11.5cm}{\caption{\label{fig:f3} 
{\sl Relation between string mass and  spectral index of CMB
anisotropy,  obtained by combining the COBE normalization of the
spectrum with the  prediction of an axionic seed model of the 
anisotropy. The dashed  lines correspond to the experimentally
allowed range of the spectral  index. The shaded area corresponds to
the theoretically expected value  of the string scale. }}}}
\end{figure}

The experimentally allowed range of the spectral index\cite{15} is, at 
present,
\beq
1\leq n \leq 1.4
\label{224}
\eeq
(I have excluded the allowed values $0.8 \leq n\leq 1$, which  
would imply in our case an 
over-critical axion production). It is remarkable that 
the corresponding allowed range of the string mass is perfectly 
compatible with  theoretical expectations\cite{16},
\beq 
0.01~\laq~ M_s/M_p~ \laq~ 0.1
\label{225}
\eeq
(see Fig. 3). Conversely, the above expected range for $M_s$ implies a 
spectral index around $1.1$ or $1.2$ (see again Fig. 3), which is also 
in very good agreement with observations. 

It must be stressed, however, that eq. (\ref{222}) is valid in the 
assumption that the inflation scale of pre-big bang models exactly 
coincides with the string mass scale. If the two scales 
were slightly different, an additional source of 
uncertainty would be introduced into the relation (\ref{222}). 

\vskip 1 cm
\renewcommand{\theequation}{3.\arabic{equation}}
\setcounter{equation}{0}
\section{Massive axions as seeds of large-scale anisotropy}
\label{sec:3}
\noindent

Up to now the discussion was devoted to massless pseudoscalar 
perturbations. It is likely, however, that axions become massive in  
the post-inflationary era: it is thus important to consider this 
possibility also. 

Let me say immediately that also in the massive case the  seed
mechanism  can work\cite{3}, and let me introduce the main
differences between the  massless and the massive case. 

A first difference is the relation between the Bardeen potential 
$\Phi$ and the 
axion energy density $\r_\sg$. In the massless case the perturbation 
equations, taking into account the important contribution of all the 
off-diagonal terms of the axion stress tensor, lead to\cite{1}
\beq
\Phi_k \sim G \left(a\over k\right)^2 \r_\sg (k).
\label{31}
\eeq
In the massive case, on the contrary, the axion stress tensor can be 
approximated as a diagonal, perfect fluid stress tensor, and we 
obtain\cite{3}
\beq
\Phi_k \sim G a^2 \eta^2 \r_\sg (k).
\label{32}
\eeq
Also, in the massless case the convolution (\ref{217}) 
for the axion energy density 
is dominated by the region\cite{1,2} $p \sim \eta^{-1}$, while in the 
massive case by\cite{1,3} $p\sim k$. In the massless case  the
integrated  SW effect is the dominant one\cite{1}, while in the
massive case the  ordinary SW effect is dominant\cite{3}. 

In spite of all these differences, the final result is similar, and in 
both cases the quadrupole coefficient is determined by the axion 
spectral energy density as 
\beq
C_2 \sim \Om_\sg^2 (k_0, \eta_0).
\label{33}
\eeq

In the massive case, however, the axion spectrum is affected by 
non-relativistic corrections. In order to include such corrections, it 
is convenient\cite{1} to distinguish between modes that become 
non-relativistic ($k/a <m$) when they are already inside the horizon 
($k/a>H$), and modes that become non-relativistic when they are still 
outside the horizon ($k/a<H$). In the first case the energy density is 
simply rescaled by the factor $m/\om$ (where $\om=k/a$ is  the
proper  frequency), and the spectrum looses a power, 
\beq
\Om_\sg \sim \om^{3-2\nu} \ra \left(m \over \om \right) \Om_\sg 
\sim \om^{2-2\nu}.
\label{34}
\eeq
In the second case, on the contrary, the spectral slope is the same as 
the relativistic one, because of the freezing of perturbations outside 
the horizon. The difference between the two regimes is graphically 
illustrated in Fig. 4, where $\om_m$ represents the limiting frequency 
of a mode that becomes non-relativistic just at the moment of 
horizon  crossing. 

\begin{figure}[htb]
\vspace{10cm}
\includegraphics{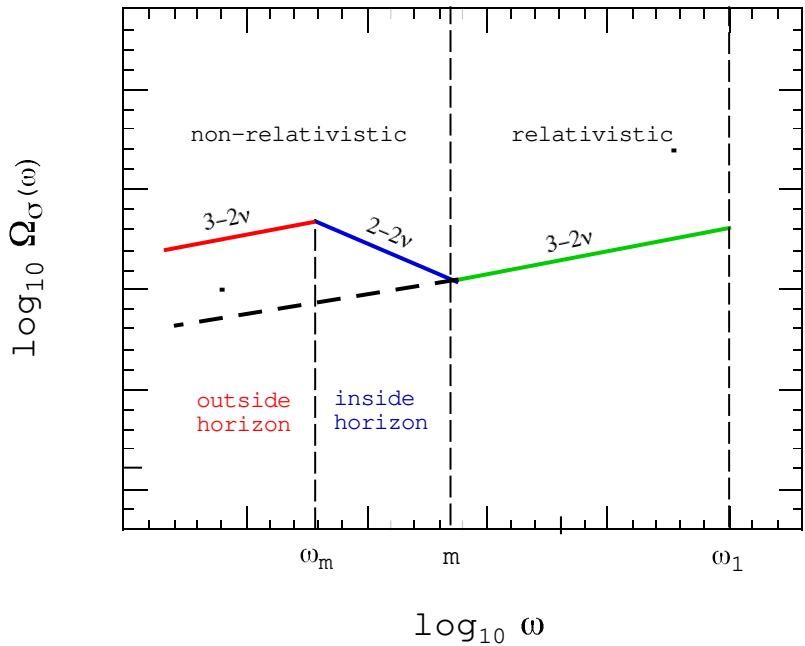}
{\caption{\label{fig:f4}
{\sl Mass-dependent enhancement of the spectrum at  low frequency,
due to  non-relativistic corrections. The non-relativistic part of the
spectrum,  to the left of $\om=m$, has a peak in correspondence of
the frequency  mode $\om_m$ that becomes non-relativistic at the
time it re-enters the  horizon. }}}
\end{figure}

As clearly shown in Fig. 4, the effect of non-relativistic corrections 
is to enhance the amplitude of the spectrum at low frequency. The 
enhancement is proportional to the square root of the axion mass  if,
at  the time when axions become massive, the scale $\om_m$ is still
outside  the horizon. In the opposite case the enhancement is linear in
the axion  mass. An explicit calculation, for a class of cosmological 
models that remain radiation-dominated from the end of inflation 
down to  the equilibrium epoch, leads in fact to the following
expression for the  low-frequency branch of the spectrum\cite{3}:
\bea
\Om_\sg &=&
\left(M_s\over M_p\right)^2 \left(m\over H_{eq}\right)^{1/2}
\left(\om\over \om_1\right)^{3-2\nu}, ~~~~~~~~~~~~
\left(m\over H_{eq}\right)^{1/2} < {T_m\over {\rm eV}}, 
\nonumber\\
&=&
\left(M_s\over M_p\right)^2 \left(m\over H_{eq}\right)
\left({\rm eV}\over T_m\right)
\left(\om\over \om_1\right)^{3-2\nu}, ~~~~~~
\left(m\over H_{eq}\right)^{1/2} > {T_m\over {\rm eV}}.
\label{35}
\eea
Here $H_{eq} \sim 10^{-27}$ eV is the Hubble scale at the time of 
matter-radiation equilibrium, and $T_m$ is the temperature scale  of
mass  generation (for instance, $T_m \sim 100$ MeV if axions become
massive at  the epoch of chiral symmetry breaking). In both cases the
slope is the  same as that of the massless spectrum, $3-2\nu$. 

The amplitude of the spectrum now depends on the axion mass, and 
the  constraint imposed by the COBE normalization (\ref{223})
necessarily  bounds the allowed range of masses. 
This might represent a problem, in general: since the slope cannot be 
too steep at low frequency (according to eq. (\ref{224})), the allowed 
mass could be too low to be compatible with realistic axion models. 

This conclusion, based on the effect illustrated in Fig. 4, 
refers however to a relativistic spectrum characterized by a constant 
slope. On the other hand, it is quite easy to imagine, and to implement 
in practice, a model of background in which the relativistic axion 
spectrum is flat enough at low frequency (as required by a fit of the 
large-scale anisotropy), and much steeper at high frequency. A simple 
example is illustrated in Fig. 5, where I have compared two 
spectra. The first one is flat everywhere, except for non-relativistic 
corrections. The second one is flat at low frequency, and steeper at 
high frequency. It is evident that the steeper and the longer the 
high-frequency branch of the spectrum, the larger is the  suppression
of  the amplitude at low frequency, and the larger is the axion mass
allowed  by the COBE normalization at $\om=\om_0$. 

\begin{figure}[htb]
\vspace{10cm}
\includegraphics{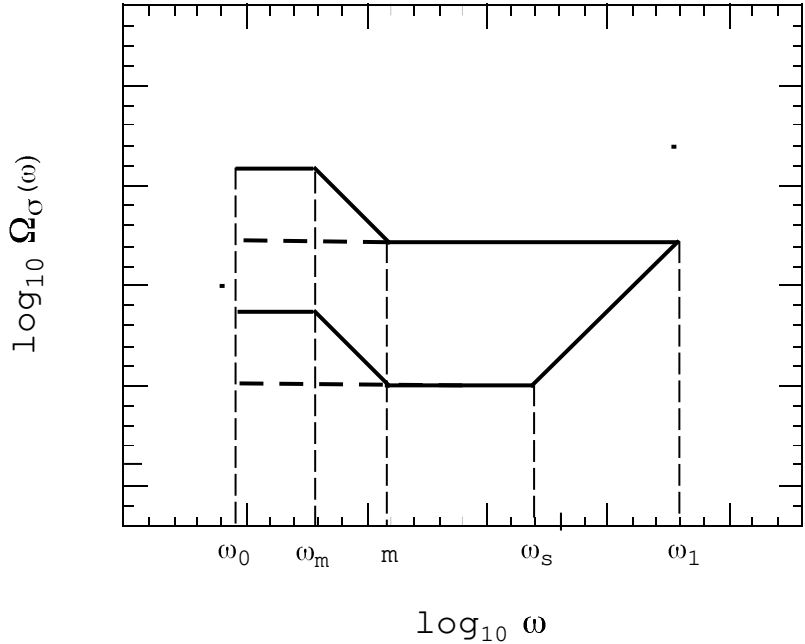}
{\caption{\label{fig:f5}
{\sl Two examples of axion spectra with non-relativistic corrections. 
Note the common normalization at the end-point frequency $\om_1$, 
in spite of the different slopes in the different frequency 
regimes. }}}
\end{figure}

We have analysed this possibility\cite{3} 
in an explicit two-parameter model of background, including exact 
solutions of the low-energy string cosmology equations with 
classical string sources. The allowed region in parameter space turns 
out to be consistent with a very wide range of axion masses, from the 
equilibrium scale $m\sim 10^{-27}$ eV up to $m \sim 100$ MeV (higher 
masses are not acceptable, because of the axion decay into photons). 
We can say, therefore, that there is no fundamental incompatibility 
between a fit of the large-scale anisotropy, and an axion mass in the 
expected range of conventional axion models. 

\vskip 1 cm
\renewcommand{\theequation}{4.\arabic{equation}}
\setcounter{equation}{0}
\section{Conclusion}
\label{sec:4}
\noindent

A stochastic cosmic background of pseudoscalar fluctuations, 
produced  with a flat enough primordial spectrum, can seed the
observed CMB  anisotropy {\em at very large angular scales}. The
end-point  normalization of the spectrum imposed by the string
cosmology scenario,  and the observational normalization at the COBE
scale, are consistent  both for massless and massive fluctuations. 

In spite of these promising results, it should be clearly stressed that 
this approach to CMB anisotropy is only the first step of a much longer 
research program, still to be implemented. Many important questions
are  still waiting for an answer, among which the crucial one, in my
opinion,  concerns the possible differences {\em at smaller angular
scales}  between this mechanism and the standard inflationary
mechanism of  anisotropy production. In particular: is the statistic
non-Gaussian? are  there shifts in the position of the Doppler peak?
etc ...

The answer to these questions is at present unclear, but we hope to 
provide answers in future papers.

\vspace{1cm}
{\it Acknowledgments:\/} I am very grateful to Ruth Durrer, Mairi 
Sakellariadou and Gabriele 
Veneziano for a fruitful collaboration which led to the 
results reported in this paper. 
I wish to thank also the Hector De Vega and Norma Sanchez for
their kind  invitation, and for the perfect organization of this
interesting Euroconference.

\end{document}